\documentclass[amsmath,amsfonts,aps,nofootinbib,preprintnumbers]{revtex4}
\usepackage{graphicx}

\begin{document}

\newcommand{\gb}[2]{{C_{#1}^{(#2)}(\eta)}}

\preprint{NT@UW-05-08}

\title{Target mass effects in deep-inelastic scattering on the
  deuteron}

\author{William Detmold} \affiliation{Department of Physics,
  University of Washington, Box 351560, Seattle, WA 98195, USA}

\begin{abstract}
  We calculate the effects of the target mass on deep-inelastic
  electron scattering on targets of spin one such as the deuteron,
  focusing on the novel structure functions that enter. It is
  important to understand these effects given the recent {\sc Hermes}
  measurements of the $b_1(x,Q^2)$ structure function of the deuteron.
  We also derive the spin-one analogues of the Wandzura-Wilczek
  relations and discuss possible calculations in lattice QCD.
\end{abstract}

\date\today \maketitle

%%%%%%%%%%%%%%%%%%%%%%%%%%%%%%%%%%%%%%%%%%%%%%%%%%%%%%%%%%%%%%%%%%%%%%
\section{Introduction}
\label{sec:introduction}
%%%%%%%%%%%%%%%%%%%%%%%%%%%%%%%%%%%%%%%%%%%%%%%%%%%%%%%%%%%%%%%%%%%%%%

The deuteron is often used as a target in deep inelastic scattering
(DIS) experiments since, neglecting nuclear effects, its
$F_{1d,2d}(x,Q^2)$ and $g_{1d,2d}(x,Q^2)$ structure functions are
simply related to those of the proton and neutron. However, the
deuteron is more interesting than this; being spin-one, it has
additional structure functions that cannot be reduced to those of the
constituent nucleons \cite{Hoodbhoy:1988am}. Recently, the {\sc
  Hermes} collaboration used a tensor polarised deuteron gas target to
obtain the first measurements of $b_{1d}(x,Q^2)$ in the range
$0.01<x<0.45$ and $0.5$~GeV$^2<Q^2<5$~GeV$^2$
\cite{Airapetian:2005cb}. Their results show it to be significantly
different from zero, rising as $x\to0$, and consistent with coherent
double scattering model predictions
\cite{Nikolaev:1996jy,Edelmann:1997ik,Bora:1997pi}.  However, these
measurements were necessarily at a relatively low $Q^2$ and power
corrections may be important in the interpretation of the data.

Corrections that scale as powers of $M^2/Q^2$ (where $M$ is the target
mass) are known as target mass corrections (TMCs) and many years ago
Nachtmann \cite{Nachtmann:1973mr} showed that it was possible to take
these into account exactly in unpolarised electron-proton DIS using
the representation theory of the Lorentz group. Georgi and Politzer
\cite{Georgi:1976vf,Georgi:1976ve} extended and clarified this, also
discussing the effects of quark masses.  More recently, target mass
(TM) effects have been investigated in the polarized nucleon
\cite{Wandzura:1977ce,Matsuda:1979ad,Piccione:1997zh,Blumlein:1998nv}
and parity violating
\cite{Georgi:1976ve,Blumlein:1998nv,Kretzer:2003iu} structure
functions, in nucleon generalised parton distributions
\cite{Geyer:2001qf,Belitsky:2001hz,Kirchner:2003wt,Geyer:2004bx} and
in the virtual photon structure functions \cite{Baba:2003ed}.  Other
power corrections arise from dynamical higher-twist effects and are
much more difficult to analyse (even their separation from QCD scaling
violations is ambiguous \cite{Mueller:1993pa}). Here we ignore these
latter terms and explore the TMCs for the leading twist structure
functions of the deuteron and other spin-one targets. We derive
spin-one analogues of the Wandzura-Wilczek \cite{Wandzura:1977qf}
relations between moments of various spin-one structure functions and
also discuss how recent developments in lattice QCD will enable
information on the underlying parton distribution functions (PDFs) in
the deuteron to be extracted directly from QCD.

%%%%%%%%%%%%%%%%%%%%%%%%%%%%%%%%%%%%%%%%%%%%%%%%%%%%%%%%%%%%%%%%%%%%%%
\section{Deep-inelastic scattering on spin-one targets}
\label{sec:deep-inel-scatt}
%%%%%%%%%%%%%%%%%%%%%%%%%%%%%%%%%%%%%%%%%%%%%%%%%%%%%%%%%%%%%%%%%%%%%%

The formalism of DIS on spin-one targets was set out by Hoodbhoy,
Jaffe and Manohar \cite{Hoodbhoy:1988am} whose nomenclature we attempt
to follow. We define the parity conserving hadronic DIS tensor and
structure functions through
\begin{eqnarray}
  \label{eq:1}
  W^{\mu\nu}(p,q,E) &\equiv& 
  \frac{1}{4\pi}\int d^4x \, e^{i q\cdot x} \langle p,E
  |\left[J^\mu(x),J^\nu(0)\right]| p, E \rangle
\\
&=& -F_1 g^{\mu\nu} + F_2 \frac{p^\mu p^\nu}{\nu} - b_1
r^{\mu\nu}+\frac{1}{6}b_2\left(s^{\mu\nu}+t^{\mu\nu}+u^{\mu\nu}\right)
+\frac{1}{2}b_3\left(s^{\mu\nu}-u^{\mu\nu}\right)
+\frac{1}{2}b_4\left(s^{\mu\nu}- t^{\mu\nu}\right)
\nonumber \\
&&\hspace*{1cm}
+\frac{i}{\nu}g_1\epsilon^{\mu\nu\lambda\sigma}q_\lambda s_\sigma
+\frac{i}{\nu^2}g_2\epsilon^{\mu\nu\lambda\sigma}q_\lambda \left(p\cdot q
  s_\sigma -s\cdot q p_\sigma\right)\,,
  \label{eq:2}
\end{eqnarray}
where $J^\mu$ is the electromagnetic current and we have ignored terms
involving $q^\mu$ or $q^\nu$ since the DIS leptonic tensor,
$\ell_{\mu\nu}$, with which $W^{\mu\nu}$ is contracted satisfies
$q^\mu\ell_{\mu\nu}=q^\nu\ell_{\mu\nu}=0$. In Eq.~(\ref{eq:2}) the
various tensors are defined as
\begin{eqnarray*}
  r^{\mu\nu}&=&\frac{1}{\nu^2}\left(q\cdot E^\ast q\cdot E -
    \frac{1}{3}\nu^2 \kappa \right) g^{\mu\nu}\,,
\end{eqnarray*}
\begin{eqnarray*}
s^{\mu\nu}&=&\frac{2}{\nu^2}\left(q\cdot E^\ast q\cdot E -
    \frac{1}{3}\nu^2 \kappa \right) \frac{p^\mu p^\nu}{\nu}\,,
\end{eqnarray*}
\begin{eqnarray*}
t^{\mu\nu}&=&\frac{1}{2\nu^2}\left( q\cdot E^\ast\left[p^\mu E^\nu+
    p^\nu E^\mu\right]+q\cdot E\left[p^\mu E^{\ast\nu}+
    p^\nu E^{\ast\mu}\right]-\frac{4}{3}\nu p^\mu p^\nu\right)\,,
\end{eqnarray*}
\begin{eqnarray*}
u^{\mu\nu}&=& \frac{1}{\nu}\left(E^{\ast\mu}E^\nu+
  E^{\ast\nu}E^\mu+\frac{2}{3}M^2g^{\mu\nu}-\frac{2}{3}p^\mu
  p^\nu\right)\,,
\end{eqnarray*}
the target polarisation is $E^\mu$ normalised such that $E^2=-M^2$ and
the generalised Pauli-Lubanski spin vector is
\begin{eqnarray*}
  s^\mu &=& -\frac{i}{M^2}\epsilon^{\mu\alpha\beta\gamma}E^\ast_\alpha
  E_\beta p_\gamma\,.
\end{eqnarray*}
Finally $\kappa=1+\frac{4x^2M^2}{Q^2}=1+\frac{M^2Q^2}{\nu^2}$ where
$\nu=p\cdot q$ and $x=Q^2/2p\cdot q$ is the Bjorken variable
($Q^2=-q^2$), $p^2=M^2$ and $p\cdot E=p\cdot E^\ast=0$.

In addition to the $F_{1,2}$ and $g_{1,2}$ structure functions
familiar from spin-half targets, four additional structure functions,
$b_{1,2,3,4}(x,Q^2)$, appear. In the parton model, $b_{1,2}(x)$ have
simple interpretations; defining $q^m_{\uparrow(\downarrow)}(x)$ as
the probability of finding a parton with spin up (down) along the beam
axis in a target with spin $m$ along the beam axis, $b_1(x)=b_2(x)/2x
= \frac{1}{2}[2q_\uparrow^0(x) - q_\uparrow^1(x) - q_\downarrow^1(x)]$
\cite{Hoodbhoy:1988am}. Since the accompanying tensors are symmetric
in their indices and vanish when averaged over target polarisation
($\langle E_\mu E^\ast_\nu\rangle_{\rm spin\;
  avg.}=\left[-M^2g_{\mu\nu}+p_\mu p_\nu\right]/3$), these functions
are measurable with unpolarised lepton beams but require a tensor
polarised target. Various models of the novel structure functions have
been proposed \cite{Miller89,Khan:1991qk,Khan:1993qb,Umnikov:1996qv,
  Nikolaev:1996jy,Edelmann:1997ik,Bora:1997pi} and bounds arising from
positivity are known \cite{Dmitrasinovic:1996xm,Bacchetta:2001rb}.

Using the optical theorem, the hadronic tensor above is related to the
forward Compton scattering amplitude $T^{\mu\nu}$,
$W^{\mu\nu}=\frac{1}{2\pi}\Im[T^{\mu\nu}]$. We decompose the tensor
structure of $T^{\mu\nu}$ as for $W^{\mu\nu}$ in Eq.~(\ref{eq:1}) with
$\{F_i,g_i,b_i\}\to\{\widetilde{F}_i,\widetilde{g}_i,\widetilde{b}_i\}$.
At leading twist, the spin-one Compton amplitude is given by
($\psi_f(x)$ is a quark field of flavour $f$ and electric charge
$Q_f$)
\begin{eqnarray}
  \label{eq:3}
T^{\mu\nu} =  \langle p,E |t^{\mu\nu} | p,E\rangle
&=& i\int d^4 x \, e^{i q\cdot x} \langle p,E|T\sum_f
  Q_f^2 \overline{\psi}_f\gamma^\mu \psi_f(x) \overline{\psi}_f\gamma^\nu
  \psi_f(0) |p,E\rangle + 
\left[\mu\leftrightarrow\nu,\,q\leftrightarrow-q\right]
\\
&=& -2\sum_{\substack{n=2 \\ {\rm even}}}^\infty 
{\cal W}_n^{(1)} g^{\mu\nu}\left(\frac{2}{Q^2}\right)^n q_{\mu_1}\ldots
q_{\mu_n} \langle p,E|{\cal O}^{\mu_1\ldots\mu_n}|p,E\rangle
\nonumber \\ && 
 +4 \sum_{\substack{n=2 \\ {\rm even}}}^\infty 
{\cal W}_n^{(2)}
g^{\mu}_{\mu_1}g^{\nu}_{\mu_2}\left(\frac{2}{Q^2}\right)^{n-1}
q_{\mu_3}\ldots 
q_{\mu_n} \langle p,E|{\cal O}^{\mu_1\ldots\mu_n}|p,E\rangle
\nonumber \\ && 
 +2i \sum_{\substack{n=1 \\ {\rm odd}}}^\infty 
{\cal W}_n^{(3)}
\epsilon^{\mu\nu\lambda}_{\quad\,\,\mu_1}q_\lambda
\left(\frac{2}{Q^2}\right)^{n} q_{\mu_2}\ldots  
q_{\mu_n} \langle p,E|\widetilde{\cal O}^{\mu_1\ldots\mu_n}|p,E\rangle\,,
  \label{eq:4}
\end{eqnarray}
where the ${\cal W}_n^{(i)}(Q^2,\mu^2,\alpha_s)$ are Wilson
coefficients (which are identical to those in spin-half DIS), $\mu$ is
the renormalisation/factorisation scale, and the vector and
axial-vector twist-two operators are
\begin{eqnarray}
  \label{eq:5}
  {\cal O}^{\mu_1\ldots\mu_n} &=&
\frac{1}{2}\sum_f
Q_f^2\left[\overline{\psi}_f \gamma^{\{\mu_1} (i\tensor{D})^{\mu_2}\ldots
  (i\tensor{D})^{\mu_n\}} \psi_f -{\rm tr}\right] \,,
\\
\label{eq:6}
  \widetilde{\cal O}^{\mu_1\ldots\mu_n} &=&
\frac{1}{2}\sum_f
Q_f^2\left[\overline{\psi}_f \gamma^{\{\mu_1} \gamma_5 (i\tensor{D})^{\mu_2}\ldots
  (i\tensor{D})^{\mu_n\}} \psi_f -{\rm tr}\right] \,.
\end{eqnarray}
Here $\tensor{D}^\mu=\frac{1}{2}(\roarrow{D}-\loarrow{D})^\mu$,
$\{\ldots\}$ indicates symmetrisation over the enclosed indices and
traces are subtracted such that the operators transform irreducibly
under the Lorentz group -- they have fixed spin.  The matrix elements
of these operators in a spin-one target of momentum $p$ and
polarisation $E$ are given by
\begin{eqnarray}
  \label{eq:7}
  \langle p,E|{\cal O}^{\mu_1\ldots\mu_n}|p,E\rangle &=&
  a_n \left[p^{\mu_1}\ldots p^{\mu_n} -{\rm
        tr}\right] + d_n
    \left[\left(E^{\ast\{\mu_1}E^{\mu_2} -
        \frac{1}{3}p^{\{\mu_1}p^{\mu_2}\right) p^{\mu_3}\ldots
      p^{\mu_n\}}  -{\rm  tr}\right]\,,
\\
\label{eq:8}
  \langle p,E|\widetilde{\cal O}^{\mu_1\ldots\mu_n}|p,E\rangle &=&
  r_n\left[s^{\{\mu_1}p^{\mu_2}\ldots p^{\mu_n\}} -{\rm  tr}\right]\,.
\end{eqnarray}
The parameters $a_n$, $d_n$ and $r_n$ (we have chosen a slightly
different normalisation for $r_n$ than Hoodbhoy {\it et al.}
\cite{Hoodbhoy:1988am} such that it is dimensionless) depend only on
the choice of target and the renormalisation scale. They encode all
the non-perturbative information about the structure of the target.
The novel effects in spin-one targets arise from the additional
structure proportional to $d_n$ in Eq.~(\ref{eq:7}); this vanishes
when averaged over polarisations even in the presence of the trace
subtractions.

%%%%%%%%%%%%%%%%%%%%%%%%%%%%%%%%%%%%%%%%%%%%%%%%%%%%%%%%%%%%%%%%%%%%%%
\section{Target mass effects}
\label{sec:target-mass-effects}
%%%%%%%%%%%%%%%%%%%%%%%%%%%%%%%%%%%%%%%%%%%%%%%%%%%%%%%%%%%%%%%%%%%%%%

Away from the Bjorken limit, the traces in Eqs.~(\ref{eq:7}) and
(\ref{eq:8}) can not be ignored, but by including them we can treat
TMCs exactly \cite{Georgi:1976vf,Georgi:1976ve}.  The contractions of
the trace subtracted tensors required in the Compton amplitude are
given explicitly in the Appendix, Eqs.  (\ref{eq:42})--(\ref{eq:46}),
and inserting these results in Eq.~(\ref{eq:4}) leads to target mass
corrected expressions for the various functions
$\{\widetilde{F}_i,\widetilde{g}_i,\widetilde{b}_i\}$. We find
\begin{eqnarray}
  \label{eq:9}
  \widetilde{F}_1(p,q) &=& \sum_{n=2,4,\ldots}^{\infty}
  2a_n\left(\frac{2\rho}{Q^2}\right)^n \left\{{\cal W}_n^{(1)} \gb{n}{1}
-\frac{2}{n(n-1)}{\cal W}_n^{(2)}  \gb{n-2}{2}\right\}\,,
\end{eqnarray}
\begin{eqnarray}
  \label{eq:10}
  \widetilde{F}_2(p,q) &=& \sum_{n=2,4,\ldots}^{\infty}
   \frac{16{\cal W}_n^{(2)}a_n}{n(n-1)}
   \left(\frac{2\rho}{Q^2}\right)^{n-1} 
   \left\{\eta \gb{n-2}{3}\right\} \,,
\end{eqnarray}
\begin{eqnarray}
  \label{eq:11}
  \widetilde{g}_1(p,q) &=& \sum_{n=1,3,\ldots}^{\infty}
   \frac{4{\cal W}_n^{(3)}  r_n\eta}{n^2}\left(\frac{2\rho}{Q^2}\right)^{n} 
  \left\{\frac{n+1}{2}\gb{n-1}{1} +(n+3)\eta\gb{n-2}{2}+4\eta^2
    \gb{n-3}{3}\right\} \,,
\end{eqnarray}
\begin{eqnarray}
  \label{eq:12}
    \widetilde{g}_2(p,q) &=& -\sum_{n=1,3,\ldots}^{\infty}
   \frac{4{\cal W}_n^{(3)}  r_n\eta}{n^2}\left(\frac{2\rho}{Q^2}\right)^{n} 
  \left\{(n+2)\eta\gb{n-2}{2}+4\eta^2
    \gb{n-3}{3}\right\} \,,
\end{eqnarray}
\begin{eqnarray}
  \label{eq:13}
  \widetilde{b}_1(p,q) &=& \sum_{n=2,4,\ldots}^{\infty}
  \frac{16 d_n}{n(n-1)} \left(\frac{2\rho}{Q^2}\right)^n 
\left\{{\cal W}_n^{(1)} \eta^2\gb{n-2}{3}
 -6{\cal W}_n^{(2)}\frac{\eta^2}{n(n-1)}\gb{n-4}{4}\right\} \,,
\end{eqnarray}
\begin{eqnarray}
  \label{eq:14}
  \widetilde{b}_2(p,q) &=&  \sum_{n=2,4,\ldots}^{\infty}
  \frac{32{\cal W}_n^{(2)} d_n}{n^2(n-1)^2} \left(\frac{2\rho}{Q^2}\right)^{n-1} 
\left\{ 24\eta^3\gb{n-4}{5}+12\eta^2\gb{n-3}{4}+\eta\gb{n-2}{3} \right\} \,,
\end{eqnarray}
\begin{eqnarray}
  \label{eq:15}
   \widetilde{b}_3(p,q) &=&  \sum_{n=2,4,\ldots}^{\infty}
  \frac{32{\cal W}_n^{(2)} d_n}{n^2(n-1)^2}\left(\frac{2\rho}{Q^2}\right)^{n-1} 
\left\{ 8\eta^3\gb{n-4}{5}+4\eta^2\gb{n-3}{4} - 
\frac{2}{3}\eta\gb{n-2}{3} \right\} \,,
\end{eqnarray}
\begin{eqnarray}
  \label{eq:16}
    \widetilde{b}_4(p,q) &=&  \sum_{n=2,4,\ldots}^{\infty}
  \frac{32{\cal W}_n^{(2)} d_n}{n^2(n-1)^2} \left(\frac{2\rho}{Q^2}\right)^{n-1} 
\left\{
  8\eta^3\gb{n-4}{5}-8\eta^2\gb{n-3}{4}
  +\frac{1}{3}\eta\gb{n-2}{3} \right\}\,,   
\end{eqnarray}
where
\begin{equation}
  \label{eq:17}
  \rho= \frac{\sqrt{p^2 q^2}}{2}, \quad\quad \eta=\frac{p\cdot q}{\sqrt{p^2 q^2}}\,,
\end{equation}
and the various $C_n^{(i)}(\eta)$ are Gegenbauer polynomials
\cite{gegenbauer}.

The TMCs in these formulae can be removed by taking only the leading
power of $\eta$ in the curly braces to give inverse powers of
$x=Q^2/4\rho\,\eta$; in this limit the above forms reduce to those
given in Ref.~\cite{Hoodbhoy:1988am}. The absence of terms
proportional to $d_n$ in $\widetilde{F}_{1,2}(p,q)$ (which results
from non-trivial cancellations) could be expected since these
correspond to the polarisation averaged structure functions.  Also,
the Callan-Gross relation, $F_2(x,Q^2) = 2 x F_1(x,Q^2)$, and its
spin-one analogue, $b_2(x,Q^2)= 2 x b_1(x,Q^2)$
\cite{Hoodbhoy:1988am}, are broken by target mass effects even at
leading order in QCD where ${\cal W}_n^{(i)}=1$.

To determine the target mass corrected structure functions, we follow
the method of Georgi and Politzer \cite{Georgi:1976ve}.  We will
outline this explicitly for $b_1$ and give the final results for the
other structure functions.  Using the series expansion of the various
Gegenbauer polynomials \cite{gegenbauer} and changing the orders of
the subsequent summations, we can rewrite Eq.~(\ref{eq:13}) as
\begin{eqnarray}
  \widetilde{b}_1(\omega) &=& 
2\sum_{\ell=0}^\infty \omega^{2\ell+2}\sum_{j=0}^\infty
\left[
\frac{{\cal W}^{(1)}_{2\ell+2j+2}d_{2\ell+2j+2}}{(2\ell+2j+2)(2\ell+2j+1)}
+2\frac{M^2}{Q^2} \frac{{\cal
    W}^{(2)}_{2\ell+2j+4}d_{2\ell+2j+4}(2\ell+j+3)}{(2\ell+2j+4)^2(2\ell+2j+3)^2} 
\right]
\frac{(2\ell+j+2)!}{j!(2\ell)!}
 \left(\frac{M^2}{Q^2}\right)^j\,,
\nonumber \\
  \label{eq:18}
\end{eqnarray}
where $\omega=1/x$ and we have suppressed the $M^2$ and $Q^2$
dependence of $\widetilde{b}_1$ for brevity.  The coefficient of
$\omega^{n-1}$ in this expansion enables us to extract the Mellin
moments of $b_1(x)$ via
\begin{eqnarray}
  \label{eq:19}
 M_n(b_1) &=&  \int_0^1 dx x^{n-2} b_1(x) = \frac{1}{2\pi i}\oint
 d\omega \omega^{-n} \widetilde{b}_1(\omega)
\\
&\hspace*{-5mm}=&\hspace*{-3mm}
 2 \sum_{j=0}^{\infty} \frac{{\cal
  W}^{(1)}_{n-1+2j}d_{n-1+2j}}{(n-1+2j)(n-2+2j)}\frac{(n-1+j)!}{j!(n-3)!}
\left(\frac{M^2}{Q^2}\right)^j
+4\frac{M^2}{Q^2}\sum_{j=0}^{\infty} \frac{{\cal
  W}^{(1)}_{n+1+2j}d_{n+1+2j}}{(n+1+2j)^2(n+2j)^2}\frac{(n+j)!}{j!(n-3)!}
\left(\frac{M^2}{Q^2}\right)^j\,.
\nonumber
\end{eqnarray}
To proceed, we introduce the functions $F^{(i)}_d(y)$ whose moments
are given by
\begin{eqnarray}
  \label{eq:20}
  {\cal W}_n^{(i)} d_n =\int_0^1 dy y^n F^{(i)}_d(y)\,,
\end{eqnarray}
whereby it is trivial to show that
\begin{eqnarray}
  \label{eq:21}
  \frac{{\cal W}_n^{(i)} d_n}{n(n-1)}=\int_0^1 dy y^{n-2} H^{(i)}_d(y) \,,
\quad
{\rm where} 
\quad
H^{(i)}_d(y)=\int_y^1 dz\int_z^1 d z^\prime F^{(i)}_d(z^\prime)\,,
\end{eqnarray}
and
\begin{eqnarray}
  \label{eq:22}
  \frac{{\cal W}_n^{(i)} d_n}{n^2(n-1)^2}=\int_0^1 dy y^{n-2} K^{(i)}_d(y) \,,
\quad
{\rm where} 
\quad
K^{(i)}_d(y)=\int_y^1 dz\int_z^1 d z^\prime H^{(i)}_d(z^\prime)\,.
\end{eqnarray}
The $F_d^{(1)}$ and $F_d^{(2)}$ {\it etc.} differ in a known,
perturbative QCD manner determined by the Wilson coefficients.  For
use below, we also define
\begin{eqnarray}
  \label{eq:23}
  {\cal W}_n^{(i)} a_n =\int_0^1 dy y^n F^{(i)}_a(y)\, ,
\quad
\quad
H^{(i)}_a(y)=\int_y^1 dz\int_z^1 d z^\prime F^{(i)}_a(z^\prime)\,,
\end{eqnarray}
and
\begin{eqnarray}
  \label{eq:24}
  {\cal W}_n^{(i)} r_n =\int_0^1 dy y^n F^{(i)}_r(y)\, ,
\quad
\quad
H^{(i)}_r(y)=\int_y^1 \frac{dz}{z}\int_z^1 d z^\prime F^{(i)}_r(z^\prime)\,.
\end{eqnarray}
from the coefficients $a_n$ and $r_n$.  Now noting that
\begin{equation}
  \label{eq:25}
  \sum_{j=0}^\infty \frac{(n+j)!}{j!n!}z^j = \frac{1}{(1-z)^{n+1}}\,,
\end{equation}
an inverse Mellin transform of Eq.~(\ref{eq:19}) gives the target mass
corrected structure function:
\begin{eqnarray}
  \label{eq:26}
  b_1(x,Q^2,M^2) &=& \frac{1}{2\pi i}\int_{-i\infty}^{i\infty} d n\
  x^{-n+1} M_n(b_1)
\\
&\hspace*{-15mm}=&\hspace*{-8mm}
\left\{2x\frac{d^2}{dx^2}\int_0^1 dy\frac{x^2}{y^3}H_d^{(1)}(y) -
  4\frac{M^2x^2}{Q^2}\frac{d^3}{dx^3}\int_0^1 dy
  \frac{x^2\; K_d^{(2)}(y) }{y(1-M^2y^2/Q^2)}\right\} 
\frac{1}{2\pi i}\int_{-i\infty}^{i\infty}dn\left[
  \frac{y/x}{1-M^2y^2/Q^2}\right]^n \,.
\nonumber
\end{eqnarray}
Using the identity $\int_{-i\infty}^{i\infty}dn\ z^n =2\pi\,
i\,\delta(\ln(z))$, this finally leads to
\begin{eqnarray}
  \label{eq:27}
  b_1(x,Q^2,M^2)&=& 2x \frac{d^2}{dx^2}\left[
  \frac{x^2}{\xi^2 r}H^{(1)}_d(\xi)\right]
 -4\frac{M^2 x^2}{Q^2}\frac{d^3}{dx^3} \left[\frac{ x^3}{\xi r}
  K^{(2)}_d(\xi)\right]\,,
\end{eqnarray}
where
\begin{equation}
  \label{eq:28}
  \xi=\frac{2x}{1+\sqrt{1+4M^2x^2/Q^2}}
\end{equation}
is the Nachtmann variable \cite{Nachtmann:1973mr} and
$r=\sqrt{1+4M^2x^2/Q^2}$.

For the other structure functions we similarly find
\begin{eqnarray}
  \label{eq:29}
  F_1(x,Q^2,M^2)&=& 
2\frac{x}{r}F_a^{(1)}(\xi)
-\frac{4M^2x^2}{Q^2}\frac{d}{dx}\left[\frac{x}{\xi
    r}H_a^{(1)}(\xi)\right] 
\,,
\end{eqnarray}
\begin{eqnarray}
  \label{eq:30}
  F_2(x,Q^2,M^2)&=& 4 x^2 \frac{d^2}{dx^2}\left[\frac{x^2}{\xi^2 r}
    H_a^{(2)}(\xi)\right]
\,,
\end{eqnarray}
\begin{eqnarray}
  \label{eq:31}
  g_1(x,Q^2,M^2) &=& 2x\left(  \frac{d^2}{dx^2}\,x - 
    \frac{d}{dx}\right) \left[ \frac{H^{(3)}_r(\xi)}{1+M^2
      \xi^2/Q^2}\right]
\,,
\end{eqnarray}
\begin{eqnarray}
\label{eq:32}
  g_2(x,Q^2,M^2) &=& -2x \frac{d^2}{dx^2}\,x
   \left[ \frac{H^{(3)}_r(\xi)}{1+M^2 \xi^2/Q^2}\right] \,,
\end{eqnarray}
\begin{eqnarray}
  \label{eq:33}
  b_2(x,Q^2,M^2)&=& 4x^2\left(
    \frac{d^4}{dx^4}x^2-4\frac{d^3}{dx^3}x
    +2\frac{d^2}{dx^2}\right)
    \left[\frac{x^2}{\xi^2 r} K_d^{(2)}(\xi)\right]
\,,
\end{eqnarray}
\begin{eqnarray}
  \label{eq:34}
  b_3(x,Q^2,M^2)&=& \frac{4}{3}x^2\left(
    \frac{d^4}{dx^4}x^2-4\frac{d^3}{dx^3}x
    -4\frac{d^2}{dx^2}\right)
    \left[\frac{x^2}{\xi^2 r} K_d^{(2)}(\xi)\right]
\,,
\end{eqnarray}
\begin{eqnarray}
  \label{eq:35}
  b_4(x,Q^2,M^2)&=& \frac{4}{3}x^2\left(
    \frac{d^4}{dx^4}x^2+8\frac{d^3}{dx^3}x
    +2\frac{d^2}{dx^2}\right)
    \left[\frac{x^2}{\xi^2 r} K_d^{(2)}(\xi)\right]
\,.
\end{eqnarray}
Since terms proportional to $d_n$ vanish identically in the $F_i$
structure functions, the form of the TMCs in these (and the $g_i$)
structure functions is the same as those in the spin-half case.
However, the parameters $a_n$, $d_n$ and $r_n$ describing the matrix
elements in Eqs.~(\ref{eq:7}) and (\ref{eq:8}) depend on the specific
target. Setting ${\cal W}_n^{(i)}=1$, our results for
$F_{1,2}(x,Q^2,M^2)$ are equivalent to those of Georgi and Politzer
\cite{Georgi:1976ve} and those for $g_{1,2}(x,Q^2,M^2)$ agree with
Refs.~\cite{Piccione:1997zh,Blumlein:1998nv} when twist-three
operators are ignored.  Since the only modifications occur because of
the additional structure in the matrix element in Eq.~(\ref{eq:7})
that is not present for spin-zero or spin-half targets, we can
immediately conclude that the target mass effects in $F_i$ and $g_i$
are of the same form in targets of any spin greater than one-half.  It
is also clear that TMCs in spin-zero targets such as the pion have the
same form as those in the unpolarised structure functions of the
proton.

Since we have only kept twist-two operators in the expansion of
Eq.~(\ref{eq:3}), Mellin moments of $g_1$ and $g_2$ above are
connected by the Wandzura-Wilczek (WW) relations
\cite{Wandzura:1977qf} even after target mass corrections are included
\cite{Piccione:1997zh,Blumlein:1998nv}. Similarly, our results for
$b_{2,3,4}$ satisfy the relations:
\begin{eqnarray}
  \label{eq:36}
  M_n(b_3)&=&\frac{1}{3}\frac{n^2-n-6}{n^2-n}M_n(b_2)\,, \\
  \label{eq:37}
  M_n(b_4)&=&\frac{1}{3}\frac{n^2-13n+24}{n^2-n}M_n(b_2)\,, 
\end{eqnarray}
where $M_n(f)\equiv\int_0^1 dx\, x^{n-2}f(x)$. These relations are
also valid in the presence of target mass and perturbative QCD effects
and can generally be expected to hold to better accuracy than the
usual WW relation since, for massless quarks, additional operators
only contribute at twist-four (as opposed to twist-three in the WW
case).

%%%%%%%%%%%%%%%%%%%%%%%%%%%%%%%%%%%%%%%%%%%%%%%%%%%%%%%%%%%%%%%%%%%%%%
\section{Discussion}
\label{sec:discussion}
%%%%%%%%%%%%%%%%%%%%%%%%%%%%%%%%%%%%%%%%%%%%%%%%%%%%%%%%%%%%%%%%%%%%%%

The results presented in the preceding section are relevant for the
data on $b_{1d}(x,Q^2)$ obtained by the {\sc Hermes} collaboration
since they work at a relatively low momentum transfer where $ 0.75 <
M_d^2/Q^2 < 6.9$. If the data is analysed to extract the twist-two
parton distributions in the deuteron ($q^{1}_{\uparrow}(x)$,
$q^{1}_{\downarrow}(x)$ and $q^{0}_{\uparrow}(x)$ in the notation of
Hoodbhoy {\it et al.}  \cite{Hoodbhoy:1988am}) using the parton model
result, $b_1(x) = \frac{1}{2}[2q_\uparrow^0(x) - q_\uparrow^1(x) -
q_\downarrow^1(x)]$, or a QCD improved form, these TM effects will
lead to significant modifications.  In order to assess the impact of
the TMCs, we first compare the effects in $F_2(x,Q^2)$ and
$g_1(x,Q^2)$ in the proton and the deuteron and then discuss the new
results for $b_{1d}(x,Q^2)$.

\begin{figure}[!t]
  \centering \includegraphics[width=0.49\columnwidth]{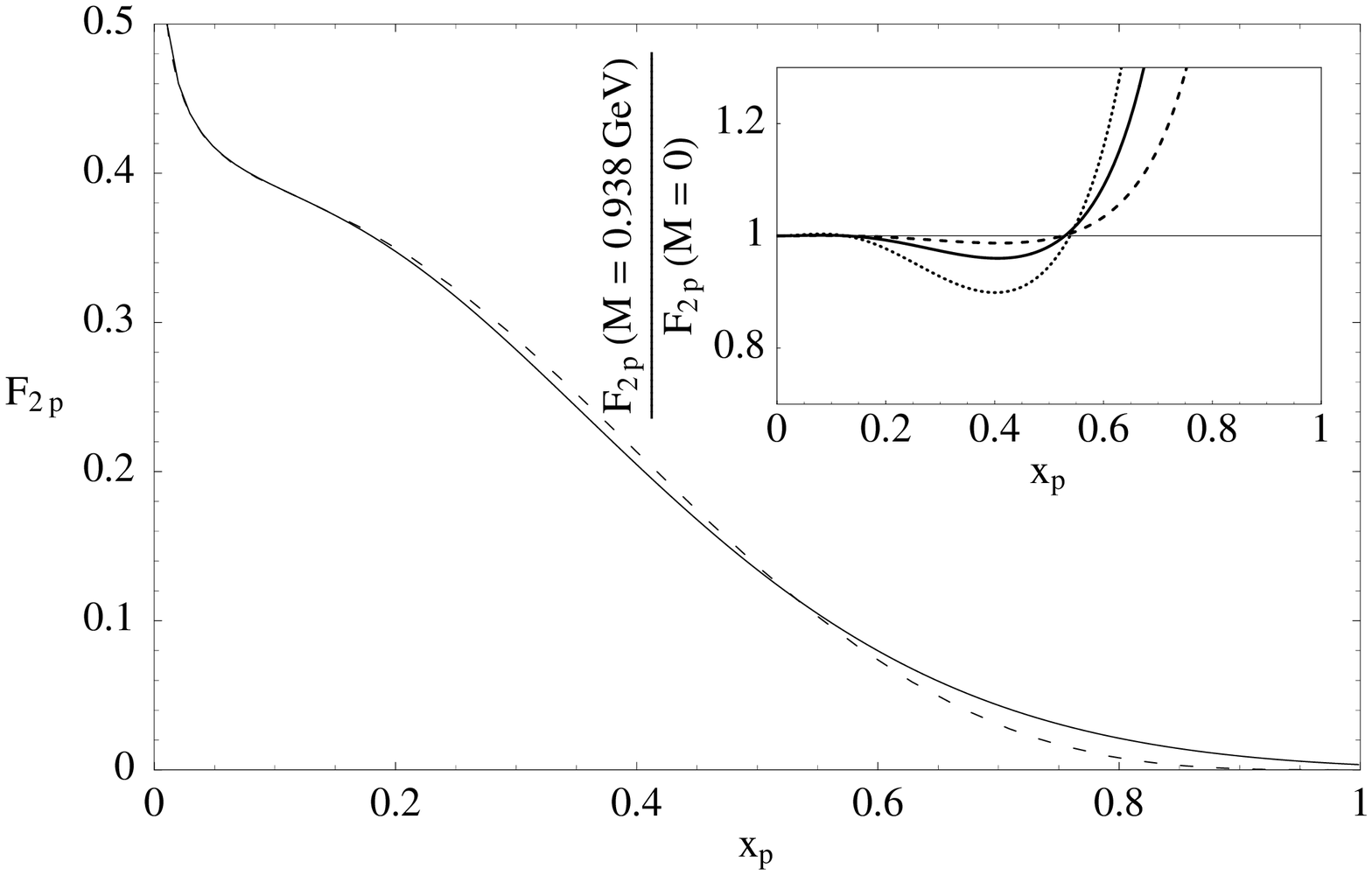}
  \hspace{1mm}\includegraphics[width=0.49\columnwidth]{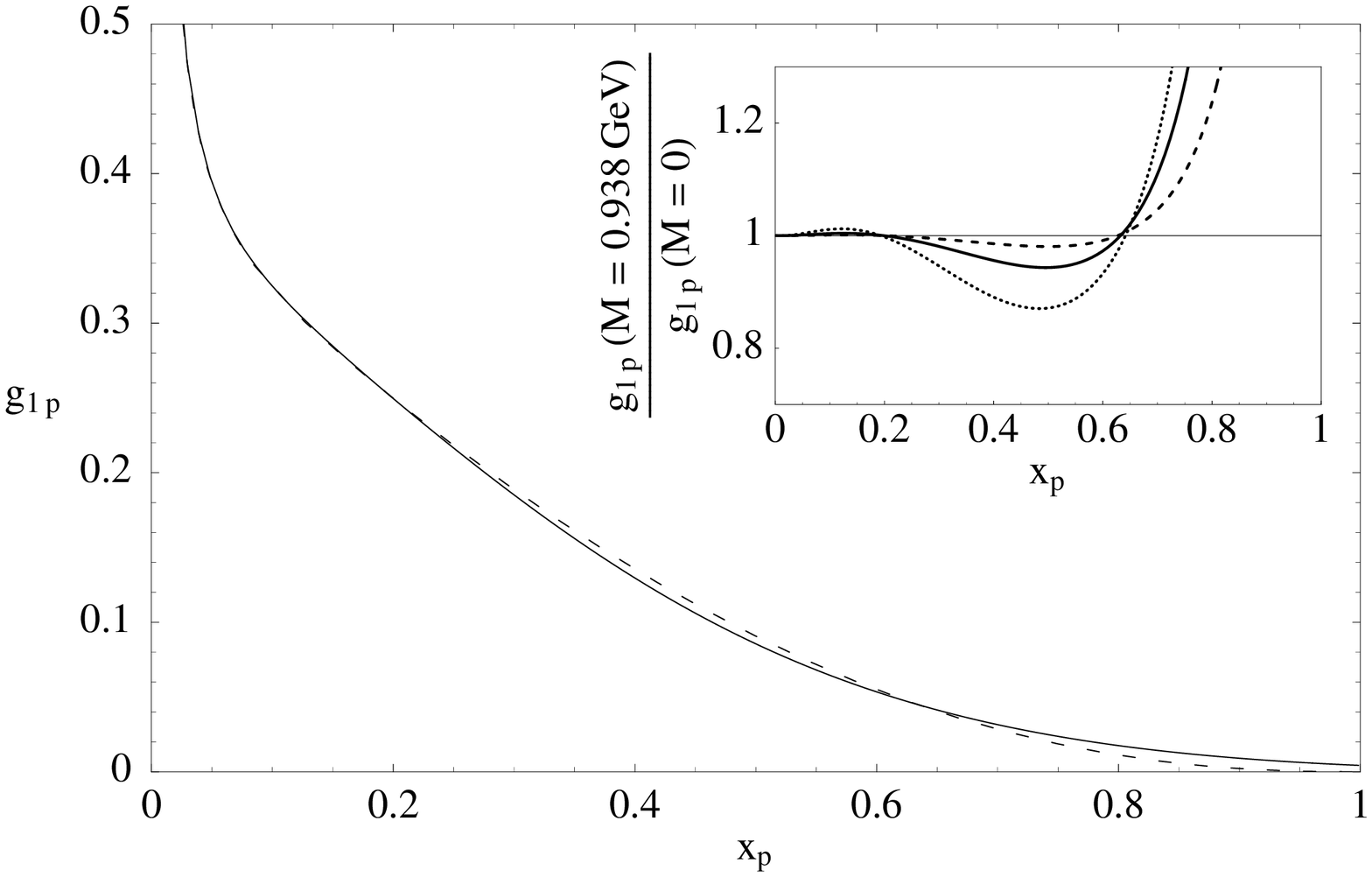}
  \caption{Structure functions of the proton. In the main panels we
    show the $F_{2p}$ (left) and $g_{1p}$ (right) structure functions
    of the proton for $M_p=0.938$~GeV (solid curve) and for $M_p=0$
    (dashed curve) at a scale $Q^2=3$~GeV$^2$. In the insets, the
    ratio of structure functions for $M_p=0.938$~GeV to those for
    $M_p=0$ is shown for $Q^2=$ 1, 3, 10 GeV$^2$ (dotted, solid and
    dashed curves, respectively).}
  \label{fig:TMCproton}
\end{figure}

Figure~\ref{fig:TMCproton} shows the $F_{2p}$ and $g_{1p}$ structure
functions of the proton at the physical mass, $M_p=0.938$~GeV, and in
the limit $M_p\to0$. To explore the TM effects in a simple manner, we
work at zeroth order in the QCD coupling. As input, we have set ${\cal
  W}_n^{(i)}=1$ and used the unpolarised parton distributions of
Alekhin \cite{Alekhin:2002fv} and the polarised PDFs of Bl\"umlein and
B\"ottcher \cite{Blumlein:2002be} (Scenario 1), both at leading order:
$F_a^{(2)}(x)=\frac{1}{9x}\left[4u(x)+d(x)+5\overline{q}(x)\right]$
and $F_r^{(3)}(x)=\frac{1}{36x}\left[4\Delta u(x)+\Delta
  d(x)+5\Delta\overline{q}(x)\right]$.  In the left panel of the
figure we show $F_{2p}(x,Q^2=3~{\rm GeV}^2)$ from Eq.~(\ref{eq:30})
evaluated with $M_p=0.938$~GeV (solid curve) and $M_p=0$ (dashed
curve). In the inset we show the ratio of these functions for $Q^2=$1,
3, 10 GeV$^2$ (dotted, solid and dashed curves, respectively). In the
right panel, the same curves are shown for the polarised structure
function $g_{1p}$ (see
Ref.~\cite{Piccione:1997zh,Blumlein:1999rv,Kretzer:2003iu} for similar
studies).  One sees the well known phenomenon that as $x\to1$, the
target mass corrected functions do not vanish as they should
kinematically. However higher twist effects that we have ignored in
the operator product expansion of Eq.~(\ref{eq:4}) become increasingly
important as $x$ approaches unity and considering only TMCs in the
large $x$ region is not reliable
\cite{DeRujula:1976tz,DeRujula:1976ih,Bitar:1978cj,Ji:1994br,Blumlein:1998nv}.

\begin{figure}[!t]
  \centering \includegraphics[width=0.49\columnwidth]{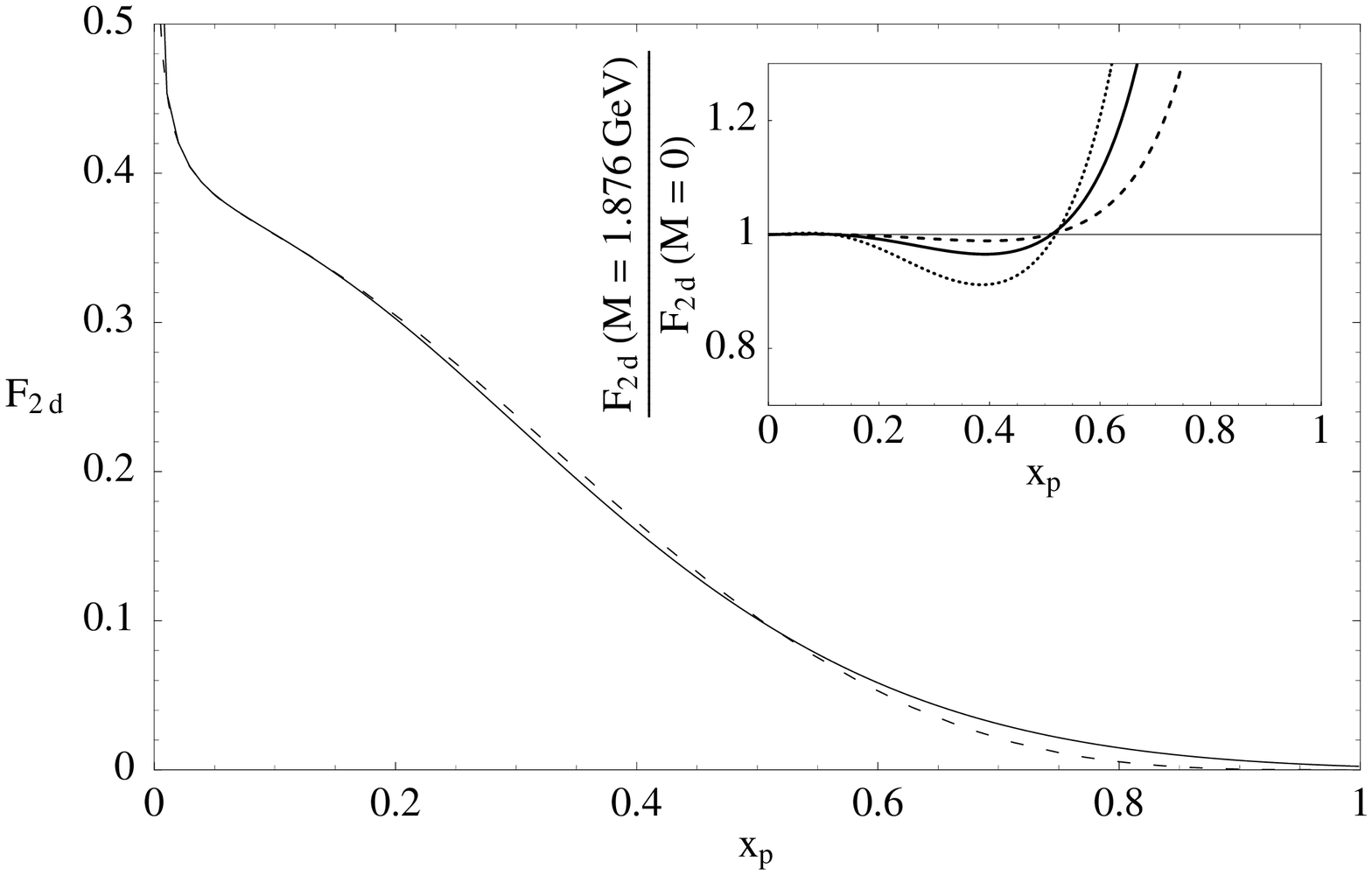}
  \hspace{1mm}\includegraphics[width=0.49\columnwidth]{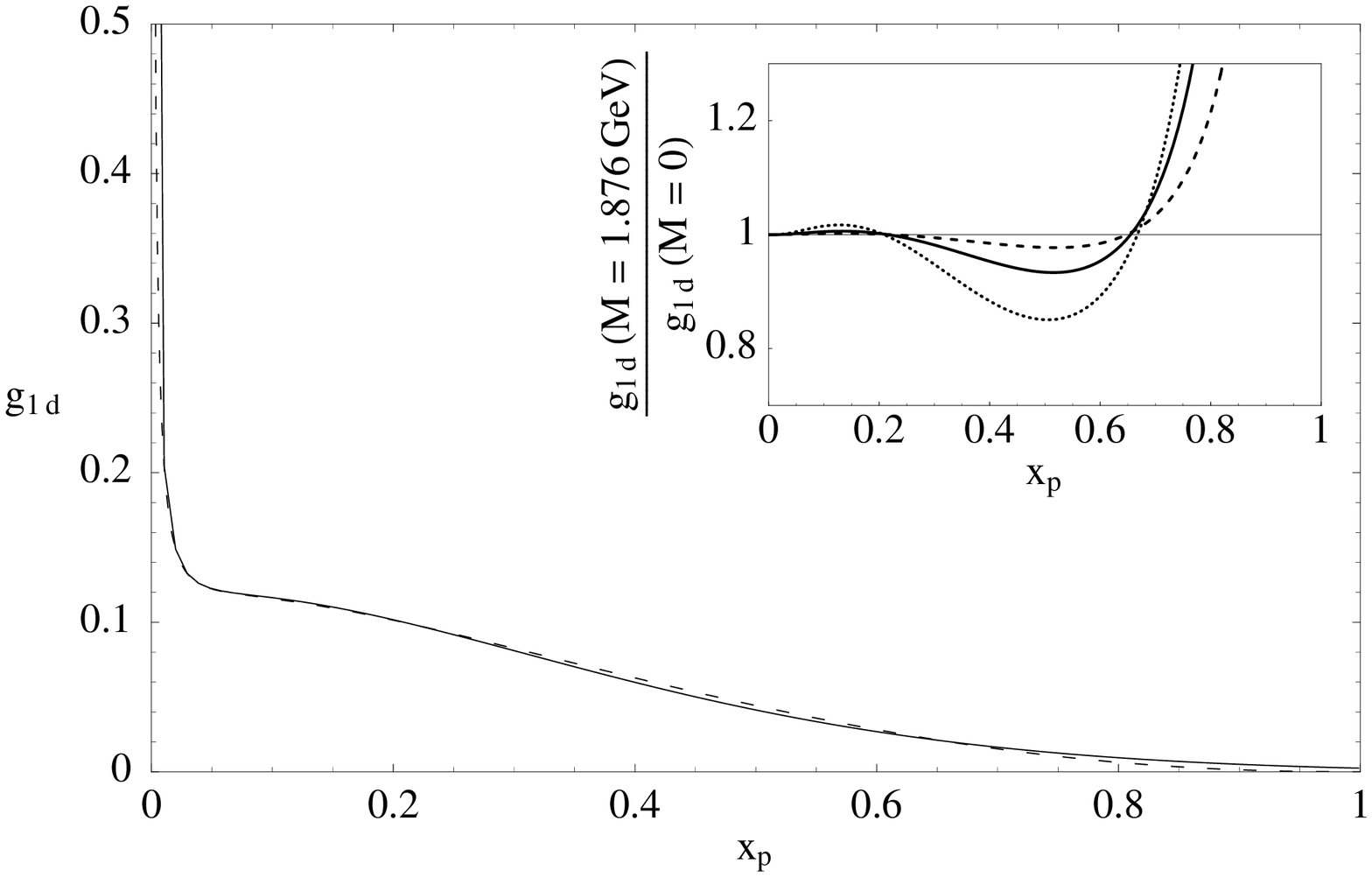}
  \caption{$F_{2d}$ (left) and $g_{1d}$ (right) structure functions of
    the deuteron. In the main figures we show the $M_d=1.876$~GeV
    (solid curve) and $M_d\to0$ (dashed curve) results at
    $Q^2=3$~GeV$^2$. In the insets the ratio of physical- to zero-
    mass structure functions is shown for $Q^2=$ 1, 3, 10 GeV$^2$
    (dotted, solid and dashed curves, respectively). }
  \label{fig:TMCdeuteron}
\end{figure}

In Figure~\ref{fig:TMCdeuteron}, we show similar results for the
deuteron structure functions $F_{2d}$ and $g_{1d}$, comparing the
structure functions for $M_d=1.876$~GeV to those in the zero
target-mass limit. Again we use the PDF parameterisations of
Refs.~\cite{Alekhin:2002fv,Blumlein:2002be}\footnote{Here we have
  ignored small nuclear effects in the deuteron, assuming it to be a
  free proton and neutron. However for heavier targets of spin one and
  higher it is interesting to consider the analogue of the EMC effect.
  For the $b_1$ structure function, one may look for the deviation of
  the ratio $b_{1A}(x)/b_{1d}(x)$ from unity for a series of nuclei of
  spin-one, {\it e.g.} $^6$Li, $^{14}$N. This will be non-zero and
  will be unrelated to the EMC effect observed in $F_2$ structure
  functions.}, setting
$F_a^{(2)}(x)=\frac{5}{18x}\left[u(x)+d(x)+2\overline{q}(x)\right]$
and $F_r^{(3)}(x)=\frac{5}{72x}\left[\Delta u(x)+\Delta
  d(x)+2\Delta\overline{q}(x)\right]$.  The larger mass of the target
and the different flavour composition of the deuteron leads to
slightly larger target mass effects in this case. In all figures, the
abscissae are defined with respect to the proton mass, $x_p=Q^2/2M_p
q_0$.

\begin{figure}[!t]
  \centering \includegraphics[width=0.49\columnwidth]{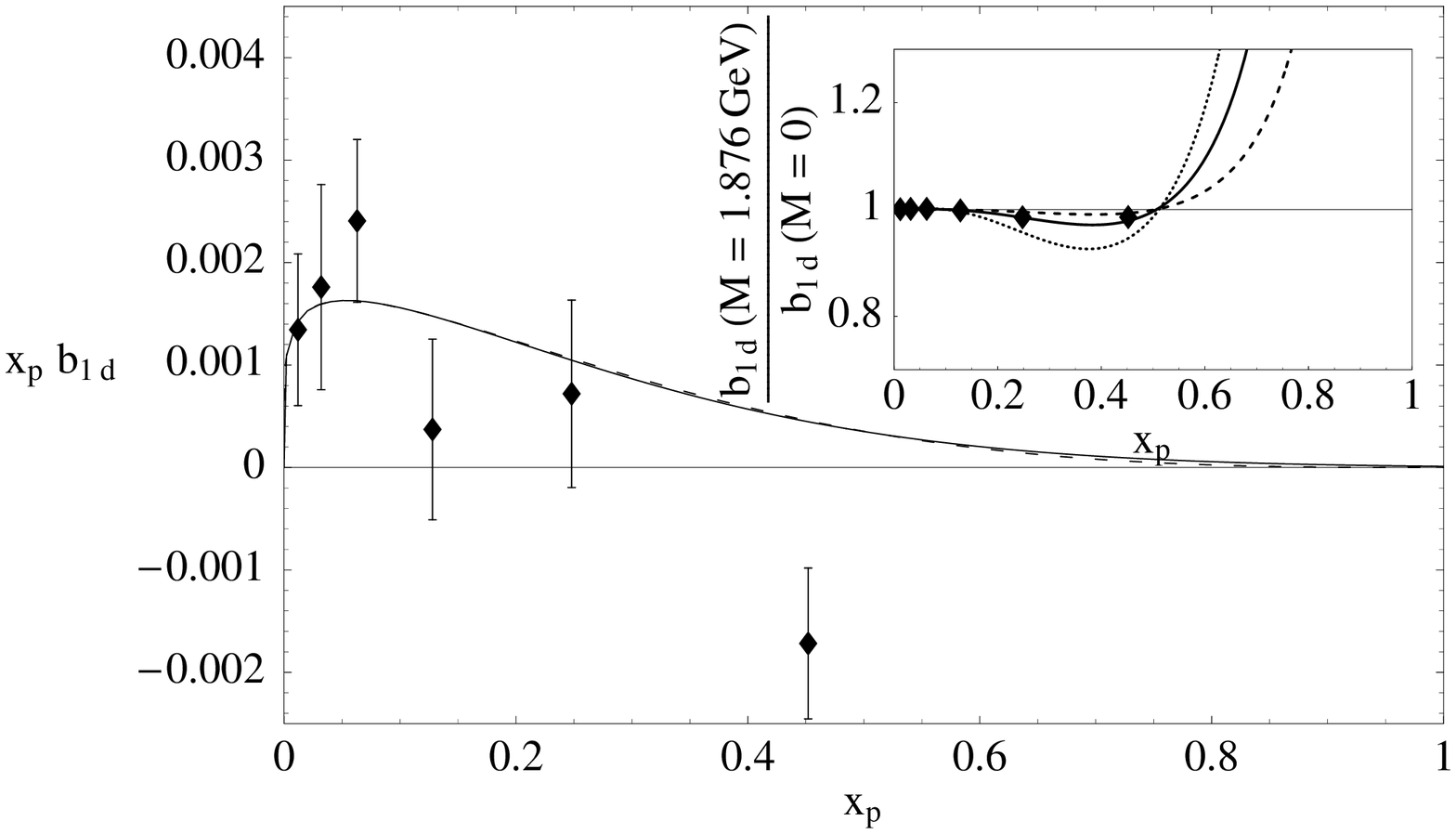}
  \hspace{1mm}\includegraphics[width=0.49\columnwidth]{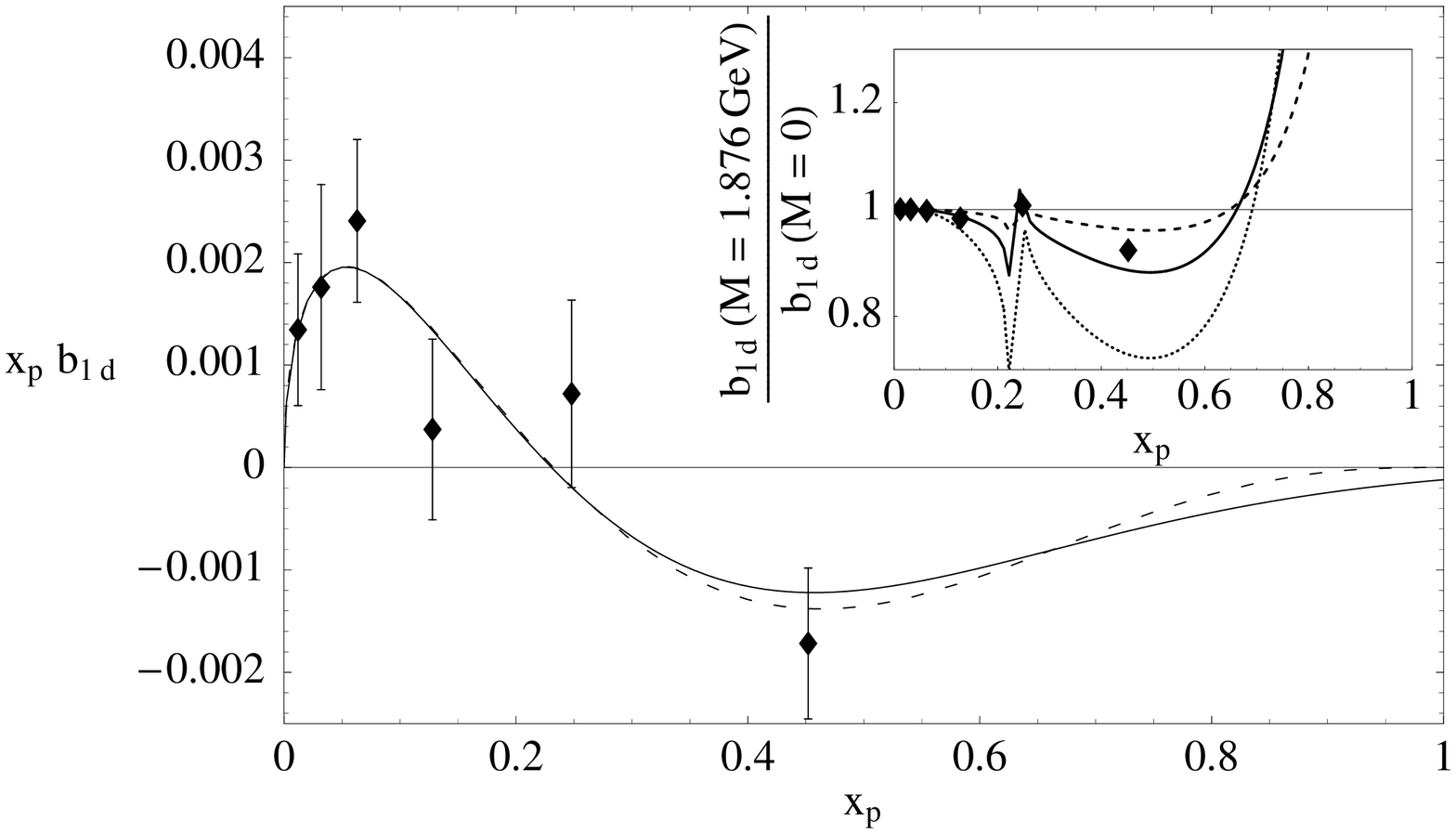}
  \caption{Calculated $b_{1d}$ structure functions of the
    deuteron. In the left plot we use the parameterisation of
    Eq.~(\ref{eq:38}) whilst the right plot uses Eq.~(\ref{eq:39}). In
    the main figure we show $x_p\,b_{1d}$ for $M_d=1.876$~GeV (solid)
    and for $M_d\to0$ (dashed) at $Q^2=3$~GeV$^2$ along with data from
    the {\sc Hermes} collaboration \protect{\cite{Airapetian:2005cb}}
    (diamonds). In the insets the ratio of physical- to zero- mass
    functions is shown for $Q^2=$ 1, 3, 10 GeV$^2$ (dotted, solid and
    dashed curves, respectively). We also display the relative shift
    at the experimental kinematics (diamonds). The kinks in the inset
    of the right plot are an artefact of parameterisation (II) having
    a zero that is shifted by the TM corrections. }
  \label{fig:TMCb1deuteron}
\end{figure}

It is less clear how to quantify the TM effects in the case of the
$b_{1d}$ structure function since we do not know the underlying
twist-two parton distributions from data at large $Q^2$. As an
estimate, we choose two forms that are fit to the data using $\chi^2$
minimisation:
\begin{eqnarray}
  \label{eq:38}
  {\rm (I)}\quad F_d^{(1)}(x)&=& A_I x^{b_I} (1-x)^3 \,,
\\
  \label{eq:39}
  {\rm (II)}\quad F_d^{(1)}(x)&=& A_{II} x^{b_{II}}
  (1-x)^{c_{II}}(1+d_{II} x) \,,
\end{eqnarray}
where the fit parameters are $A_I=0.00159$, $b_I=-1.828$ for fit (I)
and $A_{II}=0.0586$, $b_{II}=-1.532$, $c_{II}=2.86$, $d_{II}=-4.32$
for fit (II). These give a reasonable description of the data (see
dashed curves in Figure~\ref{fig:TMCb1deuteron}).  Figure
\ref{fig:TMCb1deuteron} shows the structure functions that result from
Eq.~(\ref{eq:38}) (left) and Eq.~(\ref{eq:39}) (right) for the
physical deuteron mass (solid curve) and for $M_d\to0$ (dashed curve),
both at $Q^2=3$~GeV$^2$.  In the main figures we also show the {\sc
  Hermes} data (diamonds, error bars include statistical and
systematic errors), and in the inset we show the physical- to zero-
mass $b_{1d}$ ratio at $Q^2=$ 1, 3, 10 GeV$^2$ (dotted, solid and
dashed curves) along with the ratio evaluated for the different {\sc
  Hermes} kinematics (diamonds) \cite{Airapetian:2005cb}.

From Figure~\ref{fig:TMCb1deuteron}, we see that relative shift from
TMCs can be significant (especially if the twist-two PDFs is negative
in some range of $x$), though in the region where the structure
function itself is small. If we were to go further and attempt to
extract the three independent iso-scalar twist-two PDFs in the
deuteron from data on the three structure functions $F_{2d}$, $g_{1d}$
and $b_{1d}$, it would be necessary to include these effects. However,
because of the low scales at which the measurements of $b_{1d}$ have
been made, QCD effects (which are independent of the target) and
higher-twist power corrections\footnote{A model calculation of the
  twist-four contributions to the lowest moment of $b_1$ finds a small
  effect \cite{Hoodbhoy:1990yv}.} in the data will also be important.
Without very extensive and accurate studies that would quantify such
contributions, a reliable extraction is difficult. An approach that
may help in this regard is to use lattice QCD to calculate moments of
the twist-two piece of $b_{1d}(x,Q^2)$ since matrix elements of the
twist-two operators, Eqs.~(\ref{eq:5}) and (\ref{eq:6}), can be
studied directly on the lattice. Indeed the various structure
functions of the $\rho$-meson including $b_{1\rho}$ have been explored
\cite{Best:1997qp}. For a two hadron system such as the deuteron,
calculations are more involved. Recently it has been shown that
calculations of the lattice volume dependence of two-nucleon energy
levels in background electroweak fields will lead to determinations of
electroweak properties of the deuteron \cite{Detmold:2004qn}.  Similar
methods, employing external twist-two fields, will also enable the
extraction of the matrix elements in Eqs.(\ref{eq:5}) and (\ref{eq:6})
\cite{Detmold:2004kw}.

To summarise, we have discussed target mass corrections in the
structure functions of spin-one targets. These effects are relevant
for the recent {\sc Hermes} measurements of $b_{1d}(x,Q^2)$ in the
deuteron and should be included in any extraction of the underlying
parton distributions. We have also derived non-trivial relations
between the moments of the spin-one structure functions
$b_{2,3,4}(x,Q^2)$ that are the analogues of the well-known
Wandzura-Wilczek relations. The calculations described here can easily
be extended to targets of arbitrary spin and to parity violating
structure functions. The formalism for DIS in these cases is detailed
in Refs.~\cite{Jaffe:1988up,Ravishankar:1990tz} and
\cite{Jenkins:1990dw} respectively, and the calculational procedures
are the same as those used here. However with no imminent prospect of
experimental tests we leave this to the future.

%%%%%%%%%%%%%%%%%%%%%%%%%%%%%%%%%%%%%%%%%%%%%%%%%%%%%%%%%%%%%%%%%%%%%%
\acknowledgments{The author is grateful for discussions with
  C.-J.~D.~Lin, W.~Melnitchouk, G.~A.~Miller and M.~J.~Savage. This
  work was supported by DOE contract DE-FG02-97ER41014.}
%%%%%%%%%%%%%%%%%%%%%%%%%%%%%%%%%%%%%%%%%%%%%%%%%%%%%%%%%%%%%%%%%%%%%%

%%%%%%%%%%%%%%%%%%%%%%%%%%%%%%%%%%%%%%%%%%%%%%%%%%%%%%%%%%%%%%%%%%%%%%
\appendix
\section{Traceless tensors and contractions}
%%%%%%%%%%%%%%%%%%%%%%%%%%%%%%%%%%%%%%%%%%%%%%%%%%%%%%%%%%%%%%%%%%%%%%

Let $T^{\mu_1\ldots\mu_n}=a_1^{\{\mu_1}\ldots a_n^{\mu_n\}}$ be a
completely symmetric tensor constructed of $n$ vectors $a_i^{\mu_j}$.
Then the traceless piece can be constructed as
\begin{equation}
\label{eq:40}
\mathring{T}^{\mu_1\ldots\mu_n}=\partial^{\mu_1}\ldots\partial^{\mu_n}
\mathring{T}(x)
\end{equation}
where the harmonic polynomial \cite{vilenkin93}
\begin{equation}
  \label{eq:41}
  \mathring{T}(x)=\sum_{k=0}^{\left[\frac{n}{2}\right]}
  \frac{(n-k)!}{n!k!} \left(-\frac{x^2}{4}\right)^k \square^k
  x_{\mu_1}\ldots x_{\mu_n} T^{\mu_1\ldots\mu_n}\,.
\end{equation}

This construction enables us to calculate the contractions of the
matrix elements required in Eq.~(\ref{eq:4}):
\begin{eqnarray}
  \label{eq:42}
q_{\mu_1}\ldots q_{\mu_n} \left[p^{\mu_1}\ldots p^{\mu_n} -{\rm
        tr}\right] &=&
\rho^n \gb{n}{1}\,,
\end{eqnarray}
\begin{eqnarray}
  \label{eq:43}
q_{\mu_1}\ldots q_{\mu_n} \left[E^{\ast\{\mu_1}E^{\mu_2}
  p^{\mu_3}\ldots p^{\mu_n\}} -{\rm tr}\right] &=& 
\frac{2\rho^{n-2}q^2}{n(n-1)} \left[a(q,E)\gb{n-2}{3}
- \frac{|E|^2}{4}\gb{n-2}{2}\right]\,,
\end{eqnarray}
\begin{eqnarray}
  \label{eq:44}
\epsilon^{\mu\nu\lambda}_{\quad\,\,\mu_1}q_\lambda q_{\mu_2}\ldots  
q_{\mu_n} \left[s^{\{\mu_1}p^{\mu_2}\ldots p^{\mu_n\}} -{\rm
        tr}\right] &=&
\frac{\rho^{n-1}}{n^2}\epsilon^{\mu\nu\lambda\sigma} q_\lambda \Big[
  s_\sigma\left\{ \frac{n+1}{2}\gb{n-1}{1}+ \eta\gb{n-2}{2}\right\}
\\
&&\hspace*{3cm}
+p_\sigma \frac{s\cdot q}{p\cdot q}\left\{
  (n+2)\eta\gb{n-2}{2}+4\eta^2\gb{n-3}{3}\right\}\Big] \,,
\nonumber
\end{eqnarray}
\begin{eqnarray}
  \label{eq:45}
q_{\mu_1}\ldots q_{\mu_n} \left[p^{\{\mu}p^\nu p^{\mu_1}\ldots
  p^{\mu_n\}} -{\rm tr}\right] &=&
\frac{2\rho^{n}}{(n+1)(n+2)}\Big[
  -g^{\mu\nu}\frac{\rho^2}{q^2}\gb{n}{2} + 4q^{\mu}q^{\nu}
  \frac{\rho^2}{q^4}\gb{n-2}{3}
\\
&&\hspace*{3cm} + p^{\mu}p^{\nu}
  \gb{n}{3}- 4p^{\{\mu}q^{\nu\}}
  \frac{\rho}{q^2}\gb{n-1}{3} \Big]
\nonumber\,,
\end{eqnarray}
\begin{eqnarray}
  \label{eq:46}
q_{\mu_1}\ldots q_{\mu_n} \left[E^{\ast\{\mu}E^{\nu}p^{\mu_1}\ldots
  p^{\mu_n\}} -{\rm tr}\right] &=& \frac{2\rho^{n}}{[(n+1)(n+2)]^2}
\\
&&\hspace{-5cm}
\times\Bigg[
-g^{\mu\nu}\left\{
  \frac{|E|^2}{2}\left[\gb{n}{2}-2\gb{n-2}{3}\right]
+ 6 a(q,E)\gb{n-2}{4}
\right\}
+12\frac{p^{\mu}p^{\nu}}{p^2}\left\{4a(q,E) \gb{n-2}{5}
  -\frac{|E|^2}{2}\gb{n-2}{4}\right\} 
\nonumber \\
&&\hspace{-4.4cm}
+2E^{\ast\{\mu}E^{\nu\}}\gb{n}{3}
+\frac{6}{\rho}\left(q\cdot E E^{\ast\{\mu}+q\cdot E^\ast
  E^{\{\mu}\right) p^{\nu\}}\gb{n-1}{4}
-\frac{12}{q^2}\left(q\cdot E E^{\ast\{\mu}+q\cdot E^\ast
  E^{\{\mu}\right) q^{\nu\}}\gb{n-2}{4}
\nonumber \\
&&\hspace{-4.4cm}
-2\frac{p^{\{\mu}q^{\nu\}}}{\rho} \left\{24 a(q,E) \gb{n-3}{5}
  +|E|^2\left[ \gb{n-1}{3}-3\gb{n-3}{4}\right]\right\}
\nonumber \\
&&\hspace{-4.4cm}
+2\frac{q^{\mu}q^{\nu}}{q^2} \left\{24 a(q,E) \gb{n-4}{5}
  +|E|^2\left[2 \gb{n-2}{3}-3\gb{n-4}{4}\right]\right\}
\Bigg]
\nonumber\,,
\end{eqnarray}
where $a(q,E)= \left[q\cdot E q\cdot E^\ast\right]/q^2$ and in the
last two relations we have made repeated use of the derivative
identity
\begin{equation}
  \label{eq:47}
   \frac{1}{n+1}\frac{\partial}{\partial q_\nu} q_{\mu_1}\ldots  q_{\mu_{n+1}}
  T^{\{\mu_1\ldots\mu_{n+1}\}} = q_{\mu_1}\ldots  q_{\mu_{n}}
  T^{\{\nu\mu_1\ldots\mu_{n}\}}\,.
\end{equation}

%%%%%%%%%%%%%%%%%%%%%%%%%%%%%%%%%%%%%%%%%%%%%%%%%%%%%%%%%%%%%%%%%%%%%%
\bibliography{targetmass}
%%%%%%%%%%%%%%%%%%%%%%%%%%%%%%%%%%%%%%%%%%%%%%%%%%%%%%%%%%%%%%%%%%%%%%

\end{document}